\documentclass[%
reprint,
superscriptaddress,
amsmath,amssymb,
aps,
pre,
nofootinbib
]{revtex4-2}

\usepackage{graphicx}
\usepackage{dcolumn}
\usepackage{bm}
\usepackage{mathtools}
\usepackage{amsthm}
\usepackage[hyperfootnotes=false]{hyperref}
\usepackage{float}
\usepackage{xcolor}

\allowdisplaybreaks

\newtheorem{theorem}{Theorem}[section]
\newtheorem{lemma}{Lemma}[section]
\newtheorem{proposition}{Proposition}[section]
\newtheorem{remark}{Remark}[section]

\usepackage{etoolbox}
\makeatletter
\patchcmd{\frontmatter@thanks}{;}{:}{}{}
\makeatother

\begin{document}
	
	\title{Invariant measures of exclusion processes with a look-ahead rule}
	
	\author{Lam Thi Nhung}
	\affiliation{Faculty of Information Technology, Ho Chi Minh City University of Economics and Finance, Ho Chi Minh, 700000, Vietnam}
	\author{Ngo Phuoc Nguyen Ngoc}
	\thanks{Corresponding author: ngopnguyenngoc@duytan.edu.vn}
	
	\affiliation{Institute of Research and Development, Duy Tan University, Da Nang, 550000, Vietnam.}
	\affiliation{Faculty of Natural Sciences, Duy Tan University, Da Nang 550000, Vietnam.}
	\author{Huynh Anh Thi}
	\affiliation{Institute of Research and Development, Duy Tan University, Da Nang, 550000, Vietnam.}
	\affiliation{Faculty of Natural Sciences, Duy Tan University, Da Nang 550000, Vietnam.}

	\pacs{05.40.-a, 05.70.Ln, 02.50.Ga, 89.40.Bb}

	
	\begin{abstract}
		We study a one-dimensional exclusion process with a fixed jump length $I \ge 1$ in which a particle may advance or retreat $I$ sites provided all intermediate sites are vacant, with hopping rates of Arrhenius type depending on the local headway. We identify the class of rates admitting an explicit Ising--Gibbs invariant measure, with stationarity governed by pairwise balance rather than detailed balance. In the thermodynamic limit, we derive a closed-form stationary current that recovers the mean-field prediction for look-ahead traffic-flow models exactly when particles are uncorrelated, and quantifies the correlation-induced correction otherwise.  The resulting fundamental diagram is right-skewed and non-concave, as in empirical traffic data, and for a preferred-spacing interaction--where the stationary headway distribution becomes bimodal--the mean-field approximation fails qualitatively. The closed-form current is confirmed by kinetic Monte Carlo simulations and by exact finite-size computations.
	\end{abstract}
	\maketitle
	
	\section{Introduction}
	
	The asymmetric simple exclusion process (ASEP) is one of the central models of nonequilibrium statistical mechanics, describing particles hopping on a lattice under a hard-core exclusion constraint \cite{Spitzer1970,Liggett1985,Schutz2001,MacDonald1968}. On a periodic lattice, its invariant measure is the uniform distribution over all configurations with a fixed particle number, reflecting the absence of spatial correlations in the steady state.
	
	Nontrivial stationary states arise once hopping rates depend on the local environment. A landmark example is due to Antal and Sch\"{u}tz~\cite{Antal}, who studied a next-nearest-neighbor
	interaction model in which the hopping rate of a particle is sensitive to whether the site two steps ahead is occupied. The dynamics can be summarized by the local transitions
	\begin{equation}\label{model2}
		\begin{aligned}
			&100 \longrightarrow 010 \quad \text{with rate } r,\\
			&101 \longrightarrow 011 \quad \text{with rate } q.
		\end{aligned}
	\end{equation}
	Despite the absence of detailed balance, the stationary distribution on the ring is the Ising-Gibbs measure
	\begin{equation}\label{Antal}
		\hat{\pi}(\boldsymbol{\eta})
		= \frac{1}{Z_L}
		\exp\!\Bigl\{
		-\beta J \sum_{i} \eta_i \eta_{i+1}
		- \beta h \sum_{i} \eta_i
		\Bigr\},
	\end{equation}
	where $\beta J$ is fixed by the ratio $q/r$ of hopping rates and $h$ acts as a chemical potential controlling the particle density. This striking result--a genuinely irreversible dynamics with a
	Gibbsian steady state--was subsequently extended to nearest-neighbor jumps with arbitrary finite-range interactions by Belitsky, Ngoc, and Sch\"{u}tz~\cite{Belitsky2025}.
	
	A separate motivation comes from vehicular traffic modeling. The classical Lighthill-Whitham-Richards (LWR) framework~\cite{LWR1,Whi74} leads to a symmetric, concave current $J(\rho)\propto\rho(1-\rho)$ (the so-called fundamental diagram), at odds with the asymmetric,
	non-concave shapes observed in empirical traffic data~\cite{KuG11,SCN11}. To remedy this, Sun and
	Tan~\cite{Sun2020} proposed a microscopic cellular-automata model with a \emph{look-ahead rule}: a vehicle advances $I\ge1$ cells in a single move provided all $I$ intermediate cells are empty, with
	the special case $I=2$ studied earlier in~\cite{Lee19}. Under a mean-field (propagation-of-chaos) approximation, the stationary current takes the form
	\begin{equation}\label{eq:mf_sun_tan}
		J_{\mathrm{MF}}(\rho)
		= \omega_0\,\rho\,(1-\rho)^I\,e^{-E_0},
	\end{equation}
	which is right-skewed and non-concave for $I\ge2$, in closer qualitative agreement with observations. However, the exact stationary distribution of the microscopic dynamics was not
	derived in~\cite{Sun2020,Lee19}, and the formula~\eqref{eq:mf_sun_tan} rests entirely on the mean-field assumption, leaving open the question of whether--and to what extent--inter-particle correlations modify the macroscopic flux.
	
	In this paper we address both limitations simultaneously. We study the $I$-step exclusion process ($I$-SEP), in which particles perform bidirectional jumps of fixed length $I$ provided all intermediate sites are vacant, with headway-dependent rates of Arrhenius form~\cite{Sopasakis2006}. We derive three main results. \textit{(i)}~We identify the class of rates for which the $I$-SEP admits an explicit Ising-Gibbs invariant measure on the finite ring
	(Theorem~\ref{thm:invariant_measure}), strictly generalizing \eqref{Antal} and~\cite{Belitsky2025} to arbitrary jump length $I\ge1$. Stationarity holds not by detailed balance but through a global
	pairwise balance mechanism, placing the model in the broader class of driven lattice gases with Gibbsian steady states~\cite{Antal,Belitsky2025,Katz}. \textit{(ii)}~Using an equivalence-of-ensembles argument for the headway variables (Lemma~\ref{lem:equivalence}), we derive a closed-form expression for the exact stationary current in the thermodynamic limit (Proposition~\ref{prop:current}): $J(\rho) = \rho I(r^\star-\ell^\star)e^{-\lambda(\rho)I}$, where
	$\lambda(\rho)$ is fixed by a grand-canonical density constraint. \textit{(iii)}~We show that the exact current coincides with the Sun--Tan mean-field prediction~\eqref{eq:mf_sun_tan} \emph{if and only if}
	the interaction potential is constant, i.e., precisely when particles are spatially uncorrelated. For non-trivial potentials we construct the propagation-of-chaos mean-field current $J_{\mathrm{MF}}$ of the model itself, and the ratio $J(\rho)/J_{\mathrm{MF}}(\rho)$ measures the correlation-induced correction; we illustrate it with two explicit families of interactions, for one of which (a preferred-spacing potential) the mean field fails qualitatively.  For the traffic interpretation we specialize to the totally asymmetric regime $\ell^\star=0$, where the look-ahead picture of \cite{Sun2020} applies; the bidirectional rates make the pairwise-balance structure and the equilibrium line $r^\star=\ell^\star$ transparent. We corroborate the closed-form current by direct Monte Carlo simulation of the stochastic process and by an exact finite-size canonical computation (Sec.~\ref{sec:validation}).
	
	The paper is organized as follows. Section~\ref{sec:model} defines the $I$-SEP and states the Ising-Gibbs invariant measure (Theorem~\ref{thm:invariant_measure}). Section~\ref{sec:current} derives the exact stationary current (Lemma~\ref{lem:equivalence} and Proposition~\ref{prop:current}), discusses limiting cases, compares with mean-field theory, presents two families of interaction potentials, and validates the closed-form current against kinetic Monte Carlo simulations. Section~\ref{sec:discussion} discusses implications and open problems. Proofs and computations are collected in Appendices~\ref{app:proof_thm}--\ref{app:proof_prop}.
	
	\section{Model and invariant measure}\label{sec:model}
	
	\subsection{Lattice, configurations, and dynamics}
	
	We consider a one-dimensional periodic lattice (discrete torus) $\mathbb{T}_L = \{1,\ldots,L\}$ with $N$ indistinguishable particles. Each site $x\in\mathbb{T}_L$ carries an occupation variable
	$\eta(x)\in\{0,1\}$, and a configuration is $\eta = (\eta(x))_{x\in\mathbb{T}_L}\in\{0,1\}^{\mathbb{T}_L}$.
	
	Particles perform jumps of fixed length $I\ge1$ to the right or to the left. A right jump from site $x$ to site $x+I$ is allowed only when all $I$ intermediate sites are vacant,
	\begin{equation}\label{eq:jump_right}
		\eta(x)=1,\quad \eta(x+1)=\cdots=\eta(x+I)=0,
	\end{equation}
	and analogously for a left jump from $x$ to $x-I$. All site indices are taken modulo $L$. The generator of the $I$-SEP acting on local observables $f:\{0,1\}^{\mathbb{T}_L}\to\mathbb{R}$ is
	\begin{align}
		&(\mathcal{L}f)(\eta)=\nonumber\\
		& \sum_{x\in\mathbb{T}_L}
		r_x(\eta)\,\eta(x)\prod_{j=1}^{I}(1-\eta(x+j))
		\bigl[f(\eta^{x,x+I})-f(\eta)\bigr]
		\nonumber\\
		&+
		\sum_{x\in\mathbb{T}_L}
		\ell_x(\eta)\,\eta(x)\prod_{j=1}^{I}(1-\eta(x-j))
		\bigl[f(\eta^{x,x-I})-f(\eta)\bigr],
		\label{generator}
	\end{align}
	where $\eta^{x,y}$ is the configuration obtained from $\eta$ by moving the particle from $x$ to $y$; the jump rates $r_x(\eta)$ and $\ell_x(\eta)$ are defined in Eqs.~\eqref{right_jump}--\eqref{left_jump} below.

	To describe a configuration, it is equivalent to specify it by the positions of its particles.  Label the particles by their positions $x_1<\cdots<x_N$ around the ring (with $x_{N+1}=x_1+L$) and define the \emph{headway variables}
	\begin{equation*}
		g_i := x_{i+1}-x_i, \qquad i=1,\ldots,N,
		\qquad \sum_{i=1}^{N} g_i = L.
	\end{equation*}
	 A right jump of particle $i$ requires that the destination site $x_i+I$ be vacant; since $x_{i+1}=x_i+g_i$, this site coincides with the next particle exactly when $g_i=I$. Hence a right jump of particle $i$ is admissible if and only if $g_i>I$ (equivalently $g_i\ge I+1$), and it maps $(g_{i-1},g_i)\mapsto(g_{i-1}+I,\,g_i-I)$, so that $g_i-I\ge1$ and the hard-core constraint is preserved. Symmetrically, a left jump of particle $i$ is controlled by the trailing gap $g_{i-1}$, is admissible if and only if $g_{i-1}>I$, and maps $(g_{i-1},g_i)\mapsto(g_{i-1}-I,\,g_i+I)$.
	
	We introduce an interaction potential $(J_k)_{k\ge0}$.  Throughout we adopt the convention $J_k=0$ for $k\le I$, and we assume the potential has \emph{finite range}: $J_k=0$ for $k>d$, where $d>I$ is the interaction range in headway space. The jump rates are defined by
	\begin{align}
		r_{g_i} &= r^\star\exp\!\bigl(J_{g_i-I}-J_{g_i}\bigr),
		\label{right_jump}\\
		\ell_{g_{i-1}} &= \ell^\star\exp\!\bigl(J_{g_{i-1}-I}-J_{g_{i-1}}\bigr),
		\label{left_jump}
	\end{align}
	 for $g_i>I$ (and zero otherwise). Here $r^\star,\ell^\star\ge0$ are the bare hopping rates. The quantity $J_g - J_{g-I}$ is a \emph{discrete gradient} of the interaction potential $(J_g)$ and acts as a headway-dependent energy barrier in the Arrhenius sense~\cite{Sun2020}: the right rate
	$r_g = r^\star e^{-(J_g-J_{g-I})}$ is suppressed (enhanced) when the potential increases (decreases) over a distance $I$, i.e., when $J_g - J_{g-I} > 0$ (resp.\ $<0$).

	\begin{remark}
		The rates \eqref{right_jump}--\eqref{left_jump} are of the Arrhenius form of Sun and Tan~\cite{Sun2020}, $r_g = r^\star e^{-E_b(g)}$, with a headway-dependent barrier $E_b(g):=J_g-J_{g-I}$.  Being a discrete gradient, this barrier differs in general from the linear barrier $E_b(g)=\tfrac{d-g}{d}E_0$ of~\cite{Sun2020}: a gradient telescopes to zero, $\sum_{k\ge1}\bigl(J_{r+kI}-J_{r+(k-1)I}\bigr)=0$, for any potential vanishing at large headway, while the linear barrier is strictly positive on its range. The mean-field current appropriate to the present rates is constructed in Sec.~\ref{sec:current}. For finite-range potentials, the parameter $d$ plays the role of the look-ahead range of~\cite{Sun2020}: for $g>d+I$ one has $J_{g-I}=J_g=0$, so $r_g=r^\star$ and the rate returns to its bare value, meaning that a particle ceases to react to the leading particle once the headway exceeds the threshold $d+I$.  Conversely, the rates determine the potential, $J_g=J_{g-I}-\log(r_g/r^\star)$ with $J_k=0$ for $k\le I$, so the class \eqref{right_jump}--\eqref{left_jump} may equivalently be parametrized by its rate profile; left and right rates share the same barrier, $\ell_g/\ell^\star=r_g/r^\star$.
	\end{remark}
	
	\subsection{Ising-Gibbs invariant measure}
	The invariant measure of the $I$-SEP is of Ising-Gibbs type. Expressed in occupation variables, the (unnormalized) Gibbs weight is 
	\begin{equation}\label{inv_measure_correct}
		\pi_{N,L}(\eta)
		= \exp\!\left\{
		\sum_{i=1}^{L}
	 \sum_{n=I+1}^{d}
		J_n\,\eta_i
		\prod_{k=1}^{n-1}(1-\eta_{i+k})
		\,\eta_{i+n}
		\right\},
	\end{equation}
where the interaction term $\eta_i\prod_{k=1}^{n-1}(1-\eta_{i+k})\,\eta_{i+n}$ is nonzero only when sites $i$ and $i+n$ are both occupied and all intermediate sites $i+1,\ldots,i+n-1$ are vacant--that is, when the headway between two consecutive particles equals $n$. In this sense \eqref{inv_measure_correct} is a direct
	generalization of the Ising-Gibbs measure \eqref{Antal} of Antal and Sch\"{u}tz to jump length $I\ge1$.  Writing $\hat\pi_{N,L}=\pi_{N,L}/Z_{N,L}$ with normalization $Z_{N,L}=\sum_{\eta}\pi_{N,L}(\eta)$ (the sum running over all $N$-particle configurations), the stationarity of \eqref{inv_measure_correct} is the content of the following theorem, the proof of which is given in Appendix~\ref{app:proof_thm}.
	
	\begin{theorem}[Ising-Gibbs invariant measure]\label{thm:invariant_measure} Let the $I$-SEP be 	governed by the generator \eqref{generator} with jump rates \eqref{right_jump}--\eqref{left_jump}, with a finite-range interaction potential ($J_k=0$ for $k>d$). Then the probability measure
		\begin{equation}\label{eq:inv_meas}
			\hat{\pi}_{N,L}(\eta) = \frac{1}{Z_{N,L}}\,\pi_{N,L}(\eta)
		\end{equation}
		is invariant for the process.
	\end{theorem}
	
	\begin{remark}
		The pairwise balance relation~\eqref{pairwise_balance}, established in the proof of Theorem~\ref{thm:invariant_measure}, is weaker than detailed balance: it pairs each outgoing
		transition from $\eta$ with a \emph{different} incoming transition--one involving a shifted headway index--rather than with its time-reverse. Consequently, the dynamics is generically irreversible, and the model belongs to the class of driven lattice gases with Gibbsian stationary states \cite{Antal,Belitsky2025,Katz}.
	\end{remark}
	
	\begin{remark}
		 The Antal-Sch\"{u}tz model \cite{Antal} is recovered for $I=1$, $d=2$, $\ell^\star=0$: the only admissible interaction site is $n=2$, the rates \eqref{right_jump}--\eqref{left_jump} reproduce the transitions \eqref{model2} with $J_2$ fixed by $e^{-J_2}=q/r$, and \eqref{inv_measure_correct} reduces to \eqref{Antal} (with $h=0$). The choice $I=d=1$ leaves no admissible interaction site ($J_k=0$ for $k\le I$), so the interaction term in \eqref{inv_measure_correct} is empty and one recovers the non-interacting $1$-SEP, i.e.\ the ordinary ASEP. For $I=1$ and general $d\ge2$, the measure reduces to the invariant measure of \cite{Belitsky2025}. The present theorem thus provides a unified extension to arbitrary jump length $I\ge1$.
	\end{remark}
	
	\section{Stationary current}\label{sec:current}

	\subsection{Exact stationary current}
	
	The stationary current $J$ is defined as the mean net number of particles crossing a fixed bond per unit time under the invariant measure $\hat{\pi}_{N,L}$. Since each jump has length $I$, a
	single jump contributes $I$ to the integrated current across any bond it crosses. By the periodicity of the ring, the index shift $i\mapsto i-1$ is a bijection, so $\sum_i\ell_{g_{i-1}}=\sum_i\ell_{g_i}$. Combined with translation invariance, the current takes the form
	\begin{equation}\label{eq:J_micro}
		J = \frac{I}{L}\,\mathbb{E}_{\hat{\pi}_{N,L}}
		\!\left[\sum_{i=1}^{N}\bigl(r_{g_i}-\ell_{g_{i-1}}\bigr)\right]
		= \rho\,I\,\mathbb{E}_{\nu_N}\!\bigl[r_g-\ell_g\bigr],
	\end{equation}
	where $\rho=N/L$ and $\nu_N$ is the single-headway marginal of $\hat{\pi}_{N,L}$.
	
	To evaluate \eqref{eq:J_micro} in the thermodynamic limit $N,L\to\infty$ with $N/L\to\rho$, we replace the canonical marginal $\nu_N$ by the single-headway grand-canonical law
	\begin{equation}\label{eq:nu_lambda_body}
		\nu_\lambda(g)=\frac{e^{J_g-\lambda g}}{Z(\lambda)},\quad Z(\lambda)=\sum_{g\ge1}e^{J_g-\lambda g},\quad g\ge1,
	\end{equation}
	in which $\lambda>0$ is a chemical potential conjugate to the headway (its properties are collected in Appendix~\ref{app:proof_lem}). The justification for this replacement is the following lemma.
	
	\begin{lemma}[Equivalence of ensembles]\label{lem:equivalence}
		Let $\hat{\pi}_{N,L}$ be the canonical measure defined in \eqref{eq:inv_meas}, with $(J_k)$ of finite range. For any fixed $k \ge 1$, the joint distribution of $(g_1,\ldots,g_k)$ under $\hat{\pi}_{N,L}$ converges weakly, as
		$N,L \to \infty$ with $N/L \to \rho$, to the product measure $\nu_\lambda^{\otimes k}$, with $\nu_\lambda$ the single-headway law \eqref{eq:nu_lambda_body} and $\lambda>0$ the unique solution of
		\begin{equation}\label{eq:density_constraint_lemma}
			\mathbb{E}_{\nu_\lambda}[g] = \frac{1}{\rho}.
		\end{equation}
	\end{lemma}
	
	\begin{remark}
		As shown in Appendices~\ref{app:proof_thm} and~\ref{app:proof_lem}, the measure $\hat{\pi}_{N,L}$ has a product structure in the headway variables conditioned on $\sum_{i=1}^N g_i = L$, which coincides with the canonical ensemble of a zero-range-type 
		process with single-site weights $e^{J_g}$. The weak convergence of the $k$-marginal to $\nu_\lambda^{\otimes k}$ then follows from the standard equivalence-of-ensembles 
		theorem; see Appendix~2 of \cite{KipnisLandim1999} or Theorem~1 of \cite{Grosskinsky2003}.
	\end{remark}
	
	Taking Lemma~\ref{lem:equivalence} for granted, the exact current is given by the following proposition whose proof is given in Appendix~\ref{app:proof_prop}.
	
	\begin{proposition}[Exact stationary current]\label{prop:current}
		In the thermodynamic limit $N,L\to\infty$ with $N/L\to\rho$, the stationary particle current under $\hat{\pi}_{N,L}$ is
		\begin{equation}\label{eq:current_exact}
			J(\rho)
			= \rho\,I\,(r^\star-\ell^\star)\,e^{-\lambda(\rho)\,I}.
		\end{equation}
	\end{proposition}
	
	\subsection{Special cases and comparison with mean-field theory}
	
	We now discuss the main consequences of Proposition~\ref{prop:current}.
	
	\paragraph{Non-interacting limit and recovery of mean-field.}
	When $J_g\equiv\mathrm{const}$, the measure $\nu_\lambda$ reduces to the geometric distribution  $\nu_\lambda(g) = (1-e^{-\lambda})\,e^{-\lambda(g-1)}$ on $g\ge1$, and the density constraint \eqref{eq:density_constraint_lemma},  $\mathbb{E}_{\nu_\lambda}[g]=1/(1-e^{-\lambda})=1/\rho$, gives $e^{-\lambda}=1-\rho$. Hence $e^{-\lambda I}=(1-\rho)^I$ and
	\begin{equation}\label{eq:current_noninteracting}
		J(\rho) = \rho\,I\,(r^\star-\ell^\star)\,(1-\rho)^I
		 \,=:\,J_0(\rho),
	\end{equation}
	recovering the mean-field formula \eqref{eq:mf_sun_tan} exactly (with $\omega_0 e^{-E_0}=I(r^\star-\ell^\star)$). Thus the mean-field current is the \emph{exact} stationary current when and only when particles are uncorrelated. We refer to $J_0(\rho)$ as the \emph{non-interacting current}; it serves below as the interaction-free reference.
	
	For non-constant $J_g$, the mean-field (propagation-of-chaos) current of the $I$-SEP itself is obtained by assuming i.i.d.\ headways; the density constraint $\mathbb{E}[g]=1/\rho$ then makes each headway \emph{geometric}, $\mathbb{P}(g)=\rho(1-\rho)^{g-1}$ on $g\ge1$, giving
		\begin{equation}\label{eq:mf_model}
			J_{\mathrm{MF}}(\rho)=\rho\,I\,(r^\star-\ell^\star)\,
			\mathbb{E}_{\mathrm{geo}}\!\bigl[e^{J_{g-I}-J_g}\,\mathbf{1}_{\{g>I\}}\bigr],
		\end{equation}
		where the indicator implements the admissibility constraint under which the rates \eqref{right_jump}--\eqref{left_jump} are nonzero. This is the same propagation-of-chaos approximation underlying~\eqref{eq:mf_sun_tan}, now applied to the gradient rates of the present model. For constant $J_g$ the rate factor $e^{J_{g-I}-J_g}$ is identically $1$, so \eqref{eq:mf_model} reduces to $\rho I(r^\star-\ell^\star)\,\mathbb{P}_{\mathrm{geo}}(g>I)=\rho I(r^\star-\ell^\star)(1-\rho)^I=J_0(\rho)$, the factor $(1-\rho)^I$ being the probability of an admissible gap; this recovers~\eqref{eq:mf_sun_tan}. For non-constant $J_g$, \eqref{eq:mf_model} differs from~\eqref{eq:mf_sun_tan} and is the appropriate baseline against which the quality of the mean-field approximation should be judged: the exact current~\eqref{eq:current_exact} differs from~\eqref{eq:mf_model} precisely because the true headway law $\nu_\lambda$ is not geometric, i.e.\ because $\lambda(\rho)\neq-\log(1-\rho)$. In what follows, ``mean field'' refers to~\eqref{eq:mf_model}.
	
\begin{figure*}[!htbp]
		\centering
		\includegraphics[width=\textwidth]{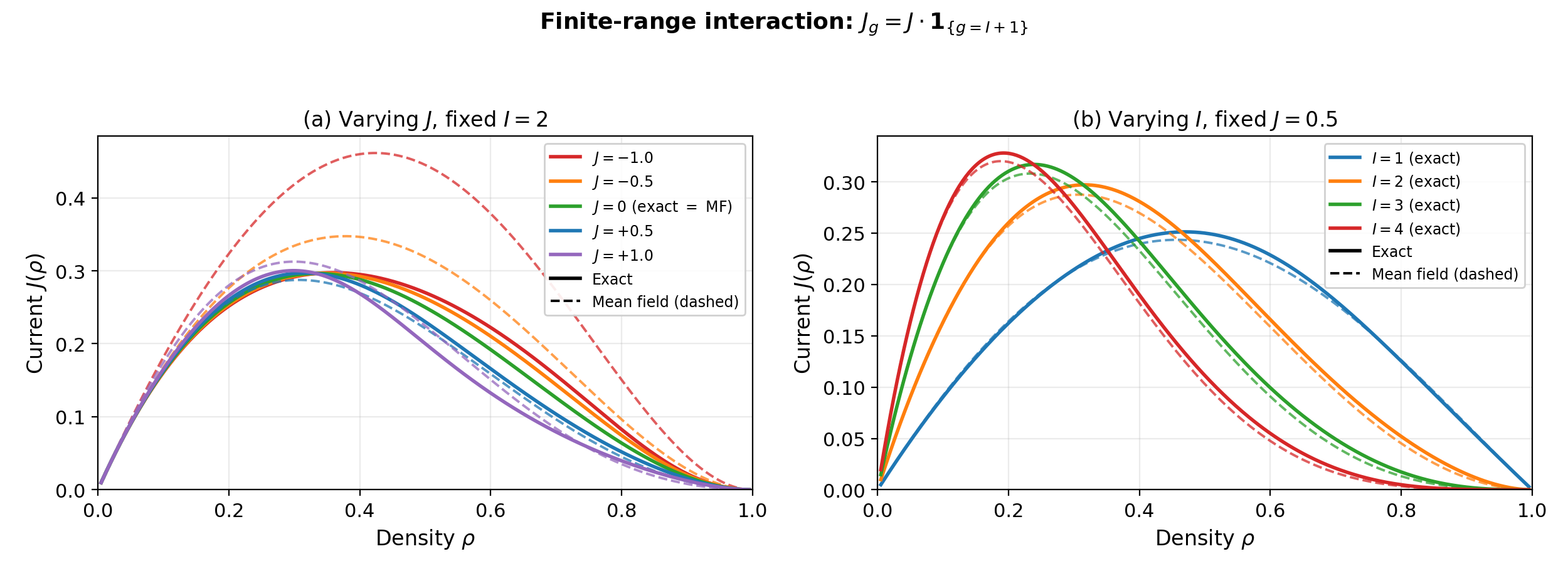}
		\caption{ Stationary current $J(\rho)$ for model
			\eqref{eq:finite_range_J}. Solid lines: exact formula
			\eqref{eq:current_exact}; dashed lines of the same color: mean field
			\eqref{eq:mf_model}. (a) Fixed $I=2$,
			$J\in\{-1,-0.5,0,0.5,1\}$; at $J=0$ exact and mean field coincide
			with the non-interacting current $J_0$. The exact peak current is
			nearly invariant in $J$, whereas the mean-field peak varies
			strongly and overshoots most severely on the repulsive side.
			(b) Fixed $J=0.5$, $I\in\{1,2,3,4\}$.}
		\label{fig:fundamental}
	\end{figure*}

	\paragraph{Effect of interactions.}
	 For non-constant $J_g$ the exact current departs from the mean-field current $J_{\mathrm{MF}}$ of Eq.~\eqref{eq:mf_model}, and the ratio $J/J_{\mathrm{MF}}$ measures the error of the propagation-of-chaos approximation. This error is quantified for explicit potentials in Sec.~\ref{sec:examples} (Figs.~\ref{fig:fundamental}--\ref{fig:Irole}); in particular, for a potential encoding a preferred spacing the mean field fails \emph{qualitatively} (Model~2 below).
	
	\paragraph{Equilibrium case.}
	If $r^\star=\ell^\star$, then $J(\rho)=0$ for all $\rho$, and the invariant measure satisfies detailed balance. The system is in thermal equilibrium independently of the interaction potential. Conversely, for $r^\star\neq\ell^\star$ the stationary current is nonzero, which is impossible for a reversible dynamics; hence the $I$-SEP is reversible if and only if $r^\star=\ell^\star$, and for $r^\star\neq\ell^\star$ it is a genuine nonequilibrium steady state, with stationarity sustained by the irreversible pairwise-balance mechanism of Theorem~\ref{thm:invariant_measure}.
	
	\paragraph{Non-concavity and inflection point.}
	The shape of the fundamental diagram is the original motivation for look-ahead models: the classical LWR flux $\rho(1-\rho)$ is concave and symmetric about $\rho=1/2$, whereas empirical traffic data are right-skewed and non-concave, with the flow maximum at a density below $1/2$~\cite{KuG11,SCN11,Sun2020}. It is therefore of interest to locate, for the present model, where the current ceases to be concave. In the non-interacting case the exact current coincides with the non-interacting form $J_0(\rho)=C\rho(1-\rho)^I$. A direct computation shows that this diagram attains its maximum at $\rho_{\mathrm{peak}}=1/(I+1)$ and, for $I\ge2$, has a unique inflection point at $\rho_c^{0}=2/(I+1)$: it is strictly concave for $I=1$, and for $I\ge2$ it is concave for $\rho<\rho_c^{0}$ and convex for $\rho>\rho_c^{0}$, so the non-concave portion occupies the high-density range $\rho>2/(I+1)$. These simple expressions are special to the non-interacting (geometric-headway) case; for non-constant $(J_g)$ both the peak and the inflection point shift away from $1/(I+1)$ and $2/(I+1)$, as we now describe.
	
	For general interactions, the inflection point is determined by the condition $J''(\rho)=0$. Using the representation $J(\rho)=\rho I e^{-I\lambda(\rho)}$, this condition can be written as
	\begin{equation}\label{eq:inflection}
		-2\lambda'(\rho)
		+ I\rho (\lambda'(\rho))^2
		- \rho \lambda''(\rho) = 0.
	\end{equation}
	Using the density constraint $\mathbb{E}_{\nu_\lambda}[g]=1/\rho$, differentiation yields
	(see Appendix~\ref{app:proof_lem})
	\[
	\lambda'(\rho) = \frac{1}{\rho^2 \sigma^2(\lambda)},
	\qquad
	\lambda''(\rho)
	= \frac{\gamma(\lambda)}{\sigma^2(\lambda)}(\lambda'(\rho))^2
	- \frac{2}{\rho^3 \sigma^2(\lambda)},
	\]
	where $\sigma^2(\lambda)$ and $\gamma(\lambda)$ denote the variance and third central moment of $\nu_\lambda$, respectively. Hence, the inflection point depends explicitly on both the second-
	and third-order moments of the headway distribution, highlighting the role of inter-particle correlations beyond mean-field theory.
	
	{\color{blue}
	 \begin{figure*}[t]
			\centering
			\includegraphics[width=\textwidth]{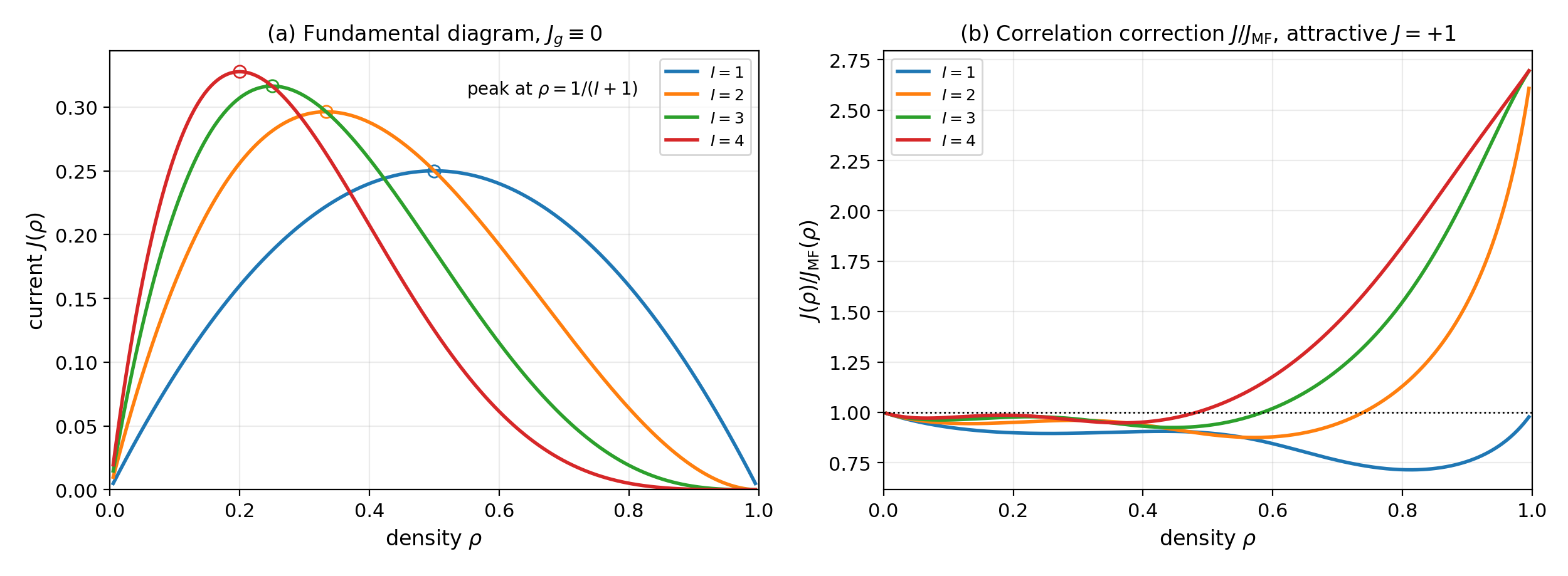}
			\caption{ Role of the jump length $I$ (totally asymmetric, $\ell^\star=0$). (a) Non-interacting fundamental diagram for $I=1,2,3,4$: increasing $I$ shifts the current maximum to $\rho=1/(I+1)$ (circles), raises its height, and makes the diagram increasingly right-skewed and non-concave.  (b) Correlation correction $J(\rho)/J_{\mathrm{MF}}(\rho)$ for an attractive interaction ($J=+1$, Model~1): the departure from unity measures the correlations missed by mean field and grows systematically with $I$, from a few tens of percent at $I=1$ to more than a factor of two at $I=4$ near jamming.}
			\label{fig:Irole}
	\end{figure*}}
	\begin{figure*}[!htbp]
		\centering
		\includegraphics[width=\textwidth]{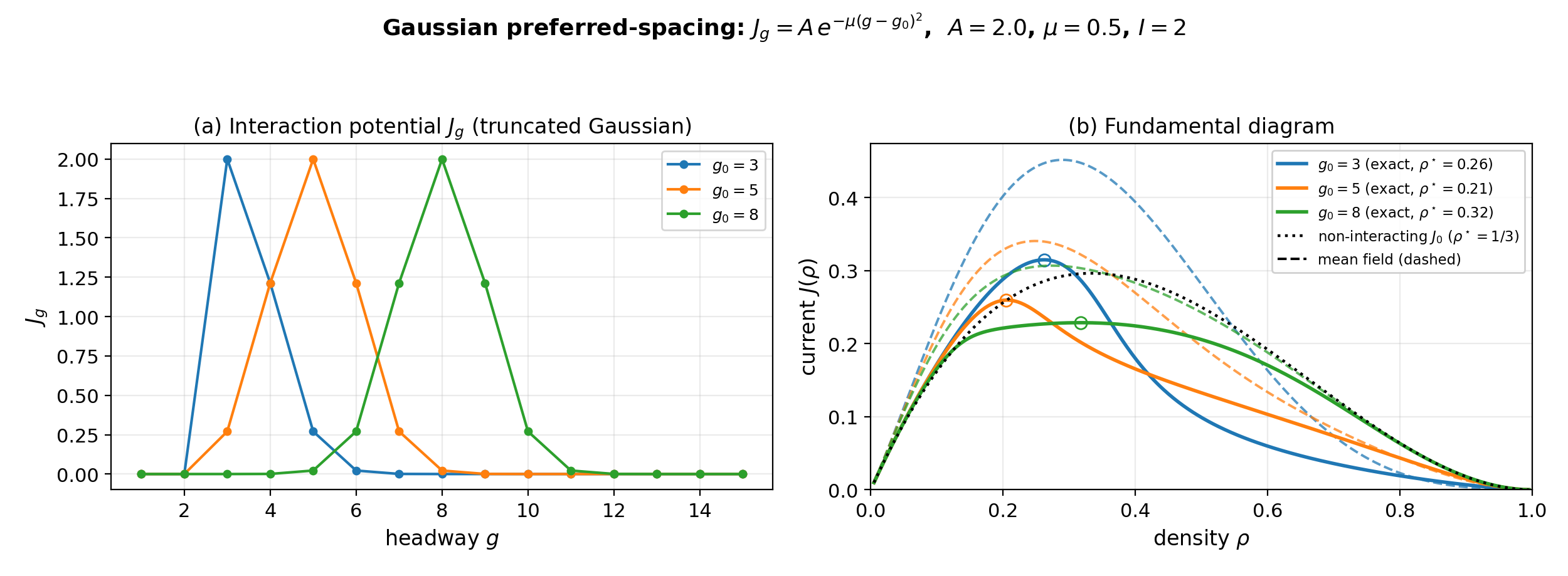}
		\caption{ Gaussian preferred-spacing model \eqref{eq:gaussian_J}
			with $I=2$, $A=2$, $\mu=0.5$, and $g_0\in\{3,5,8\}$.
			(a) Interaction potential $J_g$. (b) Exact stationary current
			$J(\rho)$ (solid) vs.\ the mean field \eqref{eq:mf_model}
			(dashed, same color) and the non-interacting current $J_0$
			(dotted black); the exact peak densities $\rho^\star$ are marked.
			Both the exact and the mean-field currents depend on $g_0$, but
			the mean field reproduces neither the location nor the magnitude
			of the exact maximum, overestimating the peak current by
			$31$--$43\%$.}
		\label{fig:gaussian}
	\end{figure*}
	
 \begin{figure*}[!htbp]
			\centering
			\includegraphics[width=\textwidth]{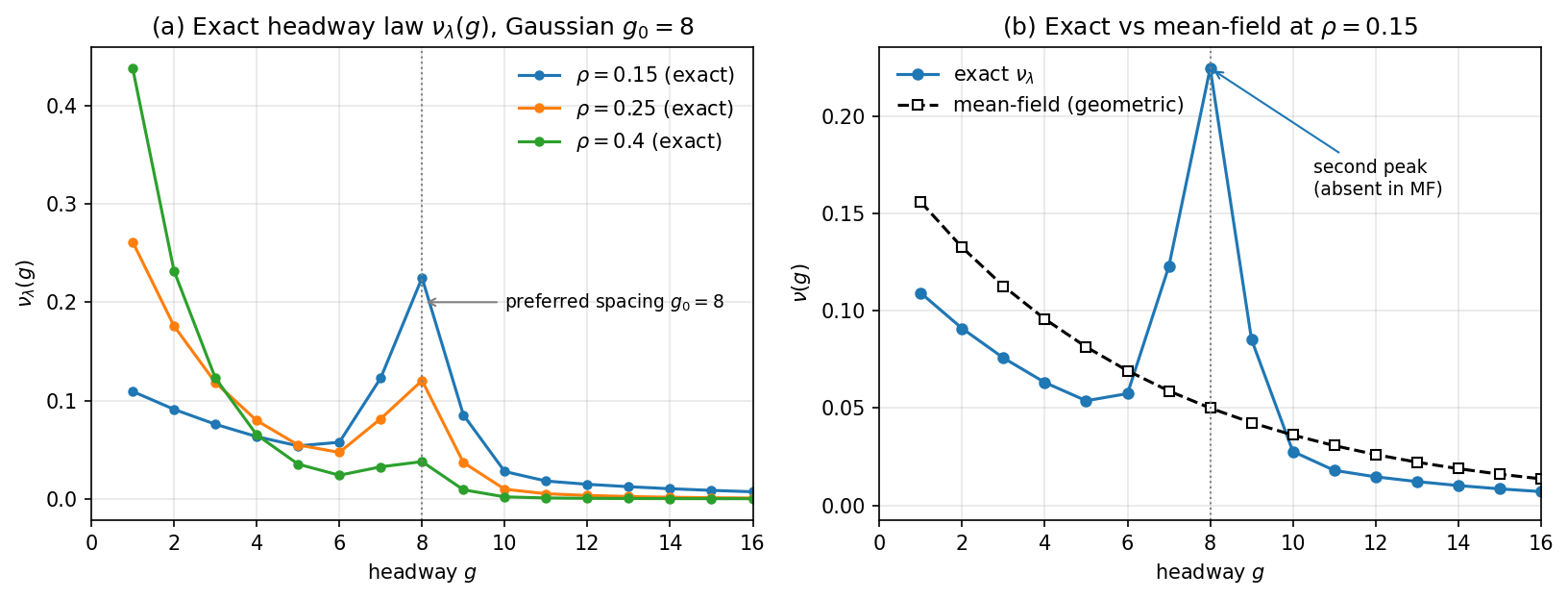}
			\caption{	 Exact stationary headway law $\nu_\lambda(g)$ for the Gaussian potential with $g_0=8$ ($A=2$, $\mu=0.5$, $I=2$). (a) For several densities the distribution is bimodal, with a peak at $g=1$ and a second peak at the preferred spacing $g_0$ (dotted line). (b) At $\rho=0.15$, the exact law (filled circles) develops a pronounced second peak at $g_0$ that is entirely absent in the geometric mean-field law (open squares).}
			\label{fig:headway}
	\end{figure*}
	
	\subsection{Examples}\label{sec:examples}
	
	We illustrate Eq.~\eqref{eq:current_exact} for two families of 	potentials. Throughout, $r^\star=1$ and $\ell^\star=0$, so $J(\rho)=\rho I\,e^{-\lambda(\rho)I}$; $\lambda(\rho)$ is found by bisection from $\mathbb{E}_{\nu_\lambda}[g]=1/\rho$.
	
	\paragraph{Model 1: finite-range interaction.}
	The minimal non-trivial extension of the Antal-Sch\"{u}tz model is 
	\begin{equation}\label{eq:finite_range_J}
		J_g = J\,\mathbf{1}_{\{g=I+1\}}, \qquad J\in\mathbb{R},
	\end{equation}
	in which the sole interaction energy arises when two consecutive particles are separated by exactly $I+1$ sites. The partition function admits the closed form
	\begin{equation}\label{eq:Z_finite_range}
		Z(\lambda)
		=  \bigl(e^{J}-1\bigr)\,e^{-(I+1)\lambda}
		+ \frac{e^{-\lambda}}{1-e^{-\lambda}},
		\qquad \lambda>0,
	\end{equation}
	where the sum runs over physical headways $g\ge1$ (with $J_g=0$ for $g\le I$).
	
	Figure~\ref{fig:fundamental} shows $J(\rho)$ for this model. Panel~(a) fixes $I=2$ and varies $J\in\{-1,-0.5,0,0.5,1\}$. At $J=0$ the exact current, the mean field \eqref{eq:mf_model}, and the non-interacting current $J_0$ all coincide, as expected from \eqref{eq:current_noninteracting}. Away from $J=0$ the exact diagram responds only weakly to the interaction: the peak current is nearly unchanged across the whole range of $J$, with only a modest shift of the peak density toward lower values as $J$ increases. The mean field, in contrast, departs from the exact curve increasingly with $|J|$, most severely on the repulsive side, where it overshoots the exact peak: it retains the Arrhenius factor $e^{J_{g-I}-J_g}$ of the rates but weighs it with geometric headway probabilities, whereas in the exact measure the relevant headways are reweighted by the Gibbs factor $e^{J_g}$, which largely compensates the rate enhancement.
	
	Panel~(b) fixes $J=0.5$ and varies $I\in\{1,2,3,4\}$. At low densities, the current is nearly insensitive to the jump length $I$, as particles are sufficiently separated and long jumps are typically allowed. In contrast, at moderate to high densities, the current decreases as $I$ increases, reflecting the reduced probability of finding an empty corridor of length $I$. The mean field (dashed) closely tracks the exact current at low and intermediate densities, the two being nearly indistinguishable there; the approximation deteriorates only at high density and large $I$, where it underestimates the exact current.

	The jump length $I$ plays a dual role, displayed in Fig.~\ref{fig:Irole}. Geometrically [panel~(a)], it sets the shape of the fundamental diagram: a jump of length $I$ requires a vacant corridor of length $I$, so increasing $I$ moves the optimal density to $\rho=1/(I+1)$ and makes the diagram increasingly right-skewed and non-concave, as discussed above. Dynamically [panel~(b)], it controls the strength of the correlations that mean-field theory discards. Because a jump samples $I$ consecutive sites at once, it probes spatial correlations on the scale $I$; the deviation of the exact current from the propagation-of-chaos prediction therefore grows systematically with $I$, from a few tens of percent at $I=1$ to more than a factor of two at $I=4$ near jamming. The single-site case $I=1$ is the least correlated, and the mean field is correspondingly most accurate there; longer jumps make the headway correlations of the interacting steady state increasingly consequential.
	
	\paragraph{Model 2: Gaussian preferred-spacing interaction.}
	 A richer example is the (truncated) Gaussian preferred-spacing potential, defined on integer headways by
	\begin{equation}\label{eq:gaussian_J}
		J_g =
		\begin{cases}
			A\exp\!\bigl(-\mu(g-g_0)^2\bigr), & I<g\le d,\\[2pt]
			0, & g>d,
		\end{cases}
	\end{equation}
with $A>0$, $\mu>0$, and $g_0>I$, which encodes a preference for the inter-particle spacing $g_0$: in traffic terms, $g_0$ is the optimal following distance, $A$ the preference strength, and $\mu^{-1/2}$ the scale of fluctuations about $g_0$.  The cutoff $d$ is chosen where the Gaussian weight is already negligible, so \eqref{eq:gaussian_J} is a genuine finite-range potential within the hypotheses of Theorem~\ref{thm:invariant_measure} and Lemma~\ref{lem:equivalence}; since the Gaussian decays superexponentially, the truncation is immaterial.\footnote{For the parameters used below ($A=2$, $\mu=0.5$, $g_0\le8$), the choice $d=16$ discards only weights $J_g<10^{-10}$ ($\sup_{g>d}J_g\approx5\times10^{-18}$), and the current changes only by terms of this order: it is indistinguishable from that of the untruncated Gaussian.}
	
	 Figure~\ref{fig:gaussian} shows results for $I=2$, $A=2$, $\mu=0.5$, and $g_0\in\{3,5,8\}$, with the cutoff $d=16$. Here the preferred-spacing well reshapes the fundamental diagram in a way the mean field misses both quantitatively and qualitatively. The optimal density itself depends on the preferred spacing non-monotonically: as $g_0$ increases through $3,5,8$, the exact peak density $\rho^\star$ first decreases and then increases ($\rho^\star\approx0.26,0.21,0.32$), a reversal driven by the competition between the hard core at $g=1$ and the preferred gap at $g_0$. The mean field cannot reproduce this trend--its peak drifts monotonically ($\rho^\star_{\mathrm{MF}}\approx0.29,0.25,0.27$)--and it overestimates the height of the maximum throughout (by $31$--$43\%$). The mechanism is the one already seen in Model~1, now amplified by the depth of the well: the mean field keeps the rate factor $e^{J_{g-I}-J_g}$ but weighs the headways geometrically, whereas the exact steady state reweights precisely those headways by the Gibbs factor $e^{J_g}$. A preferred following distance therefore leaves a clear, qualitative imprint on the macroscopic flow that propagation of chaos does not capture.
	
	 The microscopic origin of this effect is visible in the headway law itself. Figure~\ref{fig:headway} shows $\nu_\lambda(g)$ for $g_0=8$. In contrast to the geometric mean-field law, which is monotonically decreasing, the exact distribution is \emph{bimodal}: a first peak at $g=1$, reflecting the hard-core tendency of particles to sit back-to-back, and a second peak centred on the preferred spacing $g=g_0$, produced by the Gibbs weight $e^{J_g}$ favouring that headway. At $\rho=0.15$ the exact probability at $g=g_0$ exceeds the mean-field value by a factor of about $4.6$, and the probability that a headway lies in the window $g\in[g_0-1,g_0+1]$ is $43\%$, versus $15\%$ for the non-interacting (geometric) law. The steady state therefore self-organises into a mixture of two populations--tightly packed pairs and pairs at the preferred distance--separated by a depleted range of intermediate headways. This bimodal microstructure is the mechanism behind both the $g_0$-dependence of the exact current maximum and the failure of the mean-field approximation, in which headways are independent and geometrically distributed and no second peak can form.

	\begin{proposition}[Optimal jump length]\label{prop:optimal_I}
		Regard $I>0$ as a continuous parameter at fixed density $\rho$, with
		$r^\star>\ell^\star$. The current \eqref{eq:current_exact} is unimodal in $I$,
		\[
		\frac{\partial J}{\partial I}
		=\rho\,(r^\star-\ell^\star)\,e^{-\lambda(\rho)I}\bigl(1-\lambda(\rho)I\bigr),
		\]
		maximized at $I^\star(\rho)=1/\lambda(\rho)$. Since $\lambda(\rho)$ is strictly
		increasing, $I^\star(\rho)$ is strictly decreasing: a longer jump raises the
		current at low density and lowers it at high density, with
		$I^\star(\rho)\sim1/\rho=\mathbb{E}_{\nu_\lambda}[g]$ as $\rho\to0$.
	\end{proposition}
	
	\begin{remark}
		Thus transport is most efficient when the jump length matches the typical open
		corridor. In the non-interacting case the integer currents are nested,
		$J_{I+1}(\rho)>J_I(\rho)$ iff $\rho<1/(I+1)$, so consecutive diagrams in
		Fig.~\ref{fig:Irole}(a) cross exactly at the peak density $\rho=1/(I+1)$ of the
		shorter jump, and the peak current rises with $I$ to $(r^\star-\ell^\star)/e$ as $I\to\infty$.
\end{remark}
	
	\subsection{Numerical validation}\label{sec:validation}
		
		We validate the closed-form current \eqref{eq:current_exact} by two independent numerical routes. First, we perform continuous-time kinetic Monte Carlo simulations of the $I$-SEP on the ring, evolving the stochastic dynamics generated by \eqref{generator} with rates \eqref{right_jump}--\eqref{left_jump} and measuring the long-time current as $J=I\langle\text{net signed hops}\rangle/(L\,T)$. Second, we evaluate the \emph{exact} finite-$L$ canonical current by summing $\rho I\,\mathbb{E}_{\nu_N}[r_g-\ell_g]$ over the constrained measure \eqref{eq:canon_headway}; the single-headway marginal $\nu_N$ is obtained by convolution (transfer-matrix) summation over $\sum_i g_i=L$, with no sampling.
		
\begin{figure*}[t]
			\centering
			\includegraphics[width=\textwidth]{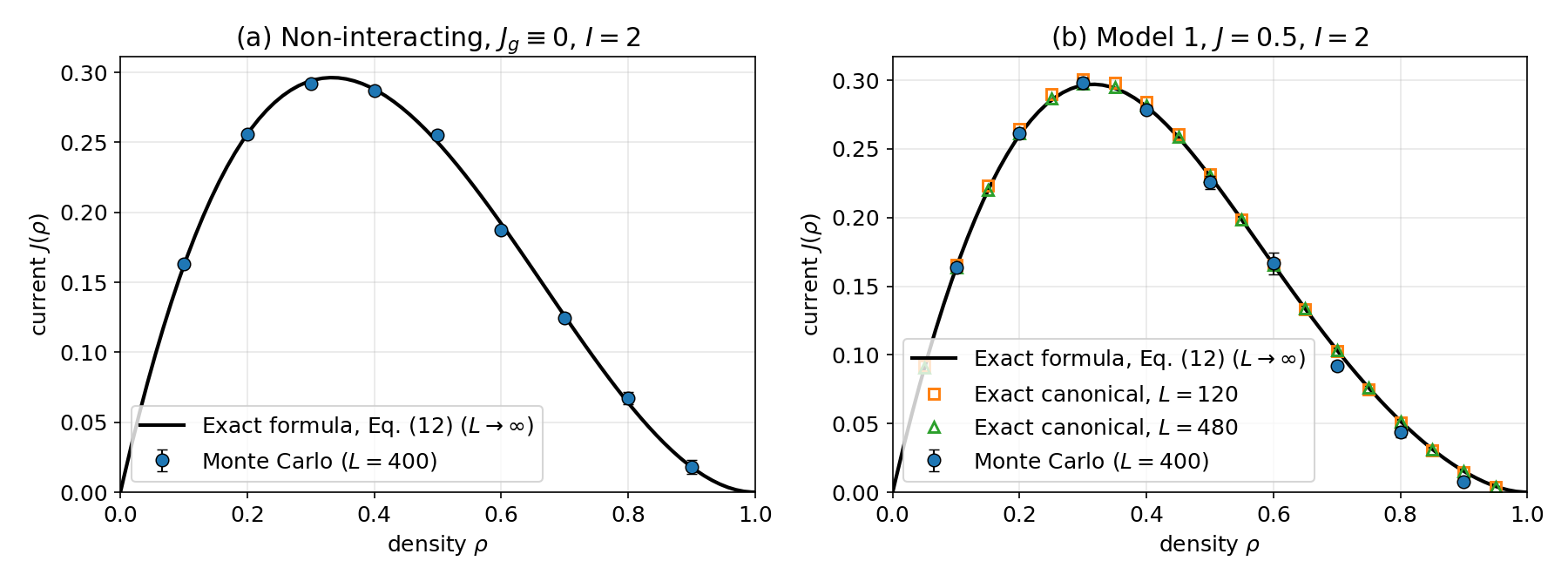}
			\caption{ Validation of the closed-form current \eqref{eq:current_exact} for $I=2$ across the full density range $\rho\in(0,1)$. (a) Non-interacting case $J_g\equiv0$: kinetic Monte Carlo (filled circles with error bars, $L=400$) against the exact formula (solid line, $L\to\infty$). (b) Model~1 with $J=0.5$: exact finite-$L$ canonical current (open symbols, $L=120,480$) and kinetic Monte Carlo (filled circles, $L=400$) against the exact formula (solid line). Agreement holds at all densities, including the high-density regime; the finite-$L$ canonical values approach the formula as $O(1/L)$.}
			\label{fig:validation}
	\end{figure*}

		Figure~\ref{fig:validation} compares both routes with \eqref{eq:current_exact} for $I=2$. In the non-interacting case [panel (a)] the simulated current agrees with the analytic curve across the whole density range, the Monte Carlo points lying on the curve to within their statistical error bars at $L=400$. For Model~1 [panel (b)] the exact canonical current converges to \eqref{eq:current_exact} as $O(1/L)$: at $\rho=0.3$ it equals $0.3008,\,0.2986,\,0.2975$ for $L=120,240,480$, approaching the formula value $0.29638$. The Monte Carlo points again lie on the analytic curve, and we verified separately that the empirical headway histogram reproduces $\nu_\lambda$.
		
	\section{Discussion}\label{sec:discussion}
	
	We have introduced and analyzed an $I$-step exclusion process ($I$-SEP) in which particles may advance or retreat by a fixed distance $I$ provided that the intermediate sites are empty.
	The hopping rates depend on the local headway through an Arrhenius-type barrier structure.
	Our main result is that, for a specific class of headway-dependent rates, the process admits an explicit invariant measure of Ising-Gibbs type, despite the absence of detailed balance.
	The stationarity mechanism relies on a global pairwise-balance structure, extending the construction of Antal and Sch\"utz~\cite{Antal} and the more recent general framework of Belitsky, Ngoc, and Sch\"utz~\cite{Belitsky2025} to exclusion processes with jumps of arbitrary length.
	
	In the thermodynamic limit we derived an exact closed-form expression for the stationary current,
	\[
	J(\rho)=\rho\,I\,(r^\star-\ell^\star)\,e^{-\lambda(\rho)I},
	\]
	where the parameter $\lambda(\rho)$ is fixed by a density constraint in the associated grand-canonical ensemble. This formula provides a direct link between the macroscopic transport properties and the statistics of the headway distribution. In particular, the mean-field stationary current proposed by Sun and Tan~\cite{Sun2020} is recovered exactly in the non-interacting case, where the headways are geometrically distributed. For nontrivial interaction potentials the deviation
	\(
	J(\rho)/J_{\mathrm{MF}}(\rho)
	\)
	quantifies the impact of inter-particle correlations on the
	macroscopic flux; we found that the propagation-of-chaos approximation, while accurate for a weak short-range potential at low and intermediate densities, fails qualitatively for a potential encoding a preferred inter-particle spacing, where the exact headway distribution becomes bimodal.
	
	Several directions for further work appear natural. First, it would be interesting to analyze hydrodynamic limits and large-scale fluctuations of the $I$-SEP, extending the macroscopic fluctuation theory developed for standard exclusion processes.  The explicit invariant measure provides exactly the input these programs require; in particular, the flux $J(\rho)$ computed here is non-concave for $I\ge2$, so the expected hydrodynamic equation $\partial_t\rho+\partial_x J(\rho)=0$ admits composite shock--rarefaction waves absent from the ordinary ASEP, a behavior that can now be predicted quantitatively from the exact fundamental diagram. Second, the present construction suggests that other classes of non-local exclusion dynamics may admit Gibbsian stationary
	states generated by pairwise-balance mechanisms. Finally, studying open systems with particle reservoirs may provide insight into boundary-induced phase transitions in look-ahead exclusion processes.
	
	\appendix
	
	\section{Proof of Theorem~\ref{thm:invariant_measure}}
	\label{app:proof_thm}
	
	We verify stationarity directly via the master equation. Since the normalization constant $Z_{N,L}$ is irrelevant for invariance, it suffices to show that the unnormalized weight $\pi_{N,L}$ satisfies
	\[
	\sum_{\eta'}\Bigl[
	\pi_{N,L}(\eta')\,\mathcal{L}(\eta',\eta)
	- \pi_{N,L}(\eta)\,\mathcal{L}(\eta,\eta')
	\Bigr] = 0,
	\]
	where $\mathcal{L}(\eta,\eta')$ denotes the transition rate from 
	configuration $\eta$ to configuration $\eta'$, read off from the 
	generator~\eqref{generator}:
	\[
	\mathcal{L}(\eta,\eta') = 
	\begin{cases}
		r_{g_i}(\eta) & \text{if } \eta' = \eta^{x_i,\,x_i+I} 
		\text{ for some } i,\\
		\ell_{g_{i-1}}(\eta) & \text{if } \eta' = \eta^{x_i,\,x_i-I} 
		\text{ for some } i,\\
		0 & \text{otherwise.}
	\end{cases}
	\]
	
	It is convenient to work with headway variables. Let $\mathbf g=(g_1,\ldots,g_N)$ be the headway configuration corresponding to $\eta$. Define the Hamiltonian
	\[
	H(\mathbf g):=\sum_{i=1}^{N} J_{g_i}.
	\]
	The candidate invariant measure \eqref{inv_measure_correct} is
	\begin{equation}\label{eq:inv_mes_head}
		\pi_{N,L}(\eta)=e^{H(\mathbf g)}.
	\end{equation}
	
	Consider a right jump of particle $i$. Such a move is allowed only if $g_i > I$. In headway variables the jump changes the pair $(g_{i-1},g_i)$ according to
	\[
	(g_{i-1},g_i) \longrightarrow (g_{i-1}+I,\, g_i-I),
	\]
	while all other headways remain unchanged.
	
	Let $\mathbf g'$ denote the configuration after the jump. The change in the Hamiltonian is therefore
	\[
	H(\mathbf g')-H(\mathbf g)
	= J_{g_{i-1}+I}+J_{g_i-I}-J_{g_{i-1}}-J_{g_i}.
	\]
	Hence the ratio of the stationary weights of the two configurations is
	\[
	\frac{\pi_{N,L}(\textbf{g}')}{\pi_{N,L}(\textbf{g})}
	= \exp\!\left(
	J_{g_{i-1}+I}+J_{g_i-I}-J_{g_{i-1}}-J_{g_i}
	\right).
	\]
	Using the rate definition \eqref{right_jump}, one gets
	\begin{align*}
		\frac{r_{g_i}}{r_{g_{i-1}+I}} = \exp\!\left(J_{g_{i-1}+I}+J_{g_i-I}-J_{g_{i-1}}-J_{g_i}\right)
	\end{align*}
	which yields the \emph{pairwise balance relation}
	\begin{equation}\label{pairwise_balance}
		\pi_{N,L}(\eta)\,r_{g_i} = \pi_{N,L}(\eta')\,r_{g_{i-1}+I},
	\end{equation}
	which asserts that the probability flux carried by the right jump of particle $i$ out of $\eta$ equals the flux of the right jump of particle $i$ into $\eta$ (from the configuration $\eta'$).	Crucially, \eqref{pairwise_balance} pairs each outgoing transition with a \emph{different} incoming transition involving a shifted headway index, rather than with its time-reverse; this is why pairwise balance is strictly weaker than detailed balance, and the dynamics remains generically irreversible.
	
	An identical relation holds for left jumps by the same argument applied to \eqref{left_jump}.  We now make the cancellation explicit. Fix a configuration $\eta$. Its outgoing transitions are the admissible right jumps (one per particle $i$ with $g_i>I$) and left jumps (one per particle with $g_{i-1}>I$). The relation \eqref{pairwise_balance} associates to the outgoing right jump of particle $i$ a unique \emph{incoming} right jump -- namely the right jump of particle $i$ from the configuration $\eta'$ obtained by the inverse map $(g_{i-1},g_i)\mapsto(g_{i-1}-I,g_i+I)$ -- of equal flux. This association $\eta\mapsto\eta'$ is a bijection between the admissible outgoing right jumps from $\eta$ and the right jumps incoming to $\eta$, because the headway pair-update $(g_{i-1},g_i)\mapsto(g_{i-1}+I,g_i-I)$ is invertible on the set of admissible pairs; the same holds for left jumps. Therefore
		\begin{align*}
			&\ \ \sum_{\eta'}\!\bigl[\pi_{N,L}(\eta')\mathcal{L}(\eta',\eta)
			-\pi_{N,L}(\eta)\mathcal{L}(\eta,\eta')\bigr]\\
			&=\sum_{i}\bigl[\pi_{N,L}(\eta^{(i,R)})\,r_{g_{i-1}+I}
			-\pi_{N,L}(\eta)\,r_{g_i}\bigr]\\
			&\quad+\sum_{i}\bigl[\pi_{N,L}(\eta^{(i,L)})\,\ell_{g_i+I}
			-\pi_{N,L}(\eta)\,\ell_{g_{i-1}}\bigr]\\
			& =0,
		\end{align*}
		by \eqref{pairwise_balance} term by term. Since this holds for every $\eta$, the master equation is satisfied. After normalization, $\hat{\pi}_{N,L}$ defined in~\eqref{eq:inv_meas} is therefore an invariant probability measure.

	\section{Grand-canonical ensemble}
	\label{app:proof_lem}
	
	 \paragraph{Canonical measure and zero-range correspondence.}
		In headway variables the invariant measure \eqref{inv_measure_correct} reads
		\begin{equation}\label{eq:canon_headway}
			\hat{\pi}_{N,L}(g_1,\ldots,g_N)
			= \frac{1}{Z_{N,L}}\prod_{i=1}^{N} w(g_i)\,
			\mathbf{1}\Bigl\{\textstyle\sum_{i=1}^N g_i=L\Bigr\},
		\end{equation}
		where $w(g):=e^{J_g}$ on $g_i\in\{1,2,\dots\}$. This is exactly the canonical measure of a zero-range process with single-site weight $w$, conditioned on the total ``mass'' $\sum_i g_i=L$. Its grand-canonical counterpart is the product measure $\bigotimes_i\nu_\lambda$, where the single-site law is defined as follows.
	
	\paragraph{Grand-canonical measure.} We collect the properties of the single-headway law $\nu_\lambda$ of \eqref{eq:nu_lambda_body},
	\begin{equation}\label{eq:nu_lambda_def}
		\nu_\lambda(g)
		= \frac{e^{J_g-\lambda g}}{Z(\lambda)},\ \
		Z(\lambda) = \sum_{g\ge1}e^{J_g-\lambda g},\ \  g\ge1,
	\end{equation}
	used in the main text. For a finite-range potential, $J_g=0$ for $g>d$, so the tail of the series is geometric and $Z(\lambda)<\infty$ for every $\lambda>0$; $\nu_\lambda$ is well defined with exponential tails. The function $\log Z(\lambda)$ is strictly convex and differentiable, with
	\[
	\frac{\mathrm{d}}{\mathrm{d}\lambda}\log Z(\lambda)
	= -\mathbb{E}_{\nu_\lambda}[g],
	\]
	so the map $\lambda\mapsto\mathbb{E}_{\nu_\lambda}[g]$ is continuous and strictly decreasing from $+\infty$ to {\color{blue} $1$}. Hence, for any prescribed density $\rho\in(0,1)$, there exists
	a unique $\lambda>0$ satisfying
	\begin{equation}\label{eq:density_constraint}
		\mathbb{E}_{\nu_\lambda}[g] = \frac{1}{\rho}.
	\end{equation}
	
	 \paragraph{Equivalence of ensembles (proof of Lemma~\ref{lem:equivalence}).}
		We show that the model falls verbatim within the zero-range setting of Gro\ss kinsky, Sch\"utz, and Spohn~\cite{Grosskinsky2003}, whose Theorem~1 then yields the claim. Under the substitution $h_i:=g_i-1\in\{0,1,2,\dots\}$, the canonical measure \eqref{eq:canon_headway} becomes the zero-range canonical measure of~\cite{Grosskinsky2003}, Eq.~(2.3), with $N$ ``sites'' (our particles), conserved ``particle number'' $M:=L-N$ (our excess headway), and single-site weight
		\[
		W(h):=e^{J_{h+1}},\qquad W(0)=e^{J_1}=1
		\]
		(recall $J_k=0$ for $k\le I$); the grand-canonical measure of~\cite{Grosskinsky2003}, Eq.~(2.6), with fugacity $\phi=e^{-\lambda}$ is then exactly our $\nu_\lambda$ shifted by one. It remains to check the hypotheses of their Theorem~1. For a finite-range potential, $W(h)=1$ for $h\ge d$, so the tail of the single-site partition function is geometric: its radius of convergence is $\phi_c=1$, and $Z(\phi)\to\infty$ as $\phi\uparrow\phi_c$, whence the grand-canonical density $R(\phi)$ increases to $R(\phi_c^-)=\infty$. The critical density of~\cite{Grosskinsky2003} is therefore infinite: every density is subcritical, no condensation occurs, and for each $\rho\in(0,1)$ there is a unique $\phi(\rho)=e^{-\lambda(\rho)}\in(0,1)$ matching $R(\phi)=\frac{1-\rho}{\rho}$, i.e., $\mathbb{E}_{\nu_\lambda}[g]=1/\rho$, which is \eqref{eq:density_constraint}. Theorem~1 of~\cite{Grosskinsky2003} (proved there by a relative-entropy argument; see also \cite{KipnisLandim1999}, Appendix~2) then gives, as $N,L\to\infty$ with $N/L\to\rho$, the pointwise convergence of every fixed-$k$ marginal of \eqref{eq:canon_headway} to $\nu_{\lambda(\rho)}^{\otimes k}$, which is Lemma~\ref{lem:equivalence}. Since the observable entering the current, $g\mapsto e^{J_{g-I}-J_g}\mathbf{1}_{\{g>I\}}$, is bounded, its canonical expectation converges to the $\nu_{\lambda(\rho)}$-expectation, the form used in Appendix~\ref{app:proof_prop}.

		\begin{remark}
			The finite range of $(J_g)$ (indeed, mere boundedness) is what places the model in the condensation-free regime of~\cite{Grosskinsky2003} at every density. For an unbounded attractive potential of the form $J_g=-b\log g$, the weight $W(h)\sim h^{-b}$ is precisely of Evans type~\cite{Evans2000}, and~\cite{Grosskinsky2003} predict a condensation transition for $b>2$: below a critical density the excess headway condenses into a single macroscopic vacant stretch, i.e., the $I$-SEP would phase-separate into a jam coexisting with an empty region. We do not pursue this here.
	\end{remark}

	\paragraph{Derivatives of $\lambda(\rho)$.}
	To locate the inflection point of the stationary current in Section~\ref{sec:current}, we compute the first and second derivatives of $\lambda(\rho)$. The parameter $\lambda(\rho)$ is determined implicitly by the density constraint
	\[
	m(\lambda) := \mathbb{E}_{\nu_\lambda}[g] = \frac{1}{\rho}.
	\]
	Differentiating both sides with respect to $\rho$ gives
	\[
	m'(\lambda)\,\lambda'(\rho) = -\frac{1}{\rho^2}.
	\]
	Since  $m(\lambda) = \sum_{g\ge 1} g\,\nu_\lambda(g)$, differentiating under the sum and using
	\[
	\frac{\mathrm{d}}{\mathrm{d}\lambda}\nu_\lambda(g)
	= \left(-g + \mathbb{E}_{\nu_\lambda}[g]\right)\nu_\lambda(g),
	\]
	which follows from $\nu_\lambda(g) = e^{J_g - \lambda g}/Z(\lambda)$ and $\frac{\mathrm{d}}{\mathrm{d}\lambda}\log Z(\lambda) = 
	-\mathbb{E}_{\nu_\lambda}[g]$, one obtains
	\begin{align*}
		m'(\lambda)
		&=  \sum_{g\ge 1} g
		\left(-g + \mathbb{E}_{\nu_\lambda}[g]\right)\nu_\lambda(g)\\
		& = -\mathbb{E}_{\nu_\lambda}[g^2]
		+ \mathbb{E}_{\nu_\lambda}[g]^2
		= -\sigma^2(\lambda) < 0.
	\end{align*}
	Hence
	\begin{equation}\label{eq:lambda_prime}
		\lambda'(\rho) = \frac{1}{\rho^2\,\sigma^2(\lambda)}.
	\end{equation}
	For the second derivative, we first compute $\frac{\mathrm{d}\sigma^2(\lambda)}{\mathrm{d}\lambda}$.
	Applying the same differentiation rule,
	\begin{align*}
		\frac{\mathrm{d}\sigma^2(\lambda)}{\mathrm{d}\lambda}
		&= \frac{\mathrm{d}}{\mathrm{d}\lambda}
		\Bigl(\mathbb{E}_{\nu_\lambda}[g^2] - m(\lambda)^2\Bigr)\\
		&= \Bigl(-\mathbb{E}_{\nu_\lambda}[g^3]
		+ \mathbb{E}_{\nu_\lambda}[g^2]\cdot m(\lambda)\Bigr)
		- 2m(\lambda)\cdot m'(\lambda)\\
		&= -\mathbb{E}_{\nu_\lambda}[g^3]
		+ 3m(\lambda)\,\mathbb{E}_{\nu_\lambda}[g^2]
		- 2m(\lambda)^3
		= -\gamma(\lambda),
	\end{align*}
	where we substituted $m'(\lambda) = -\sigma^2(\lambda)$ and recognized the result as $-\mathbb{E}_{\nu_\lambda}
	[(g-m(\lambda))^3]$, with $\gamma(\lambda) = \mathbb{E}_{\nu_\lambda}[(g-m(\lambda))^3]$ the third central moment of $\nu_\lambda$. Differentiating \eqref{eq:lambda_prime} with respect to $\rho$ and applying the chain rule then yields
	\begin{equation}\label{eq:lambda_second}
		\lambda''(\rho)
		= \frac{\gamma(\lambda)}{\sigma^2(\lambda)}
		\bigl(\lambda'(\rho)\bigr)^2
		- \frac{2}{\rho^3\,\sigma^2(\lambda)}.
	\end{equation}
	Equations~\eqref{eq:lambda_prime}--\eqref{eq:lambda_second} are used in Section~\ref{sec:current} to characterize the inflection point of the stationary current $J(\rho)$.

	\medskip
	\section{Proof of Proposition~\ref{prop:current}}
	\label{app:proof_prop}
	
	Since each successful jump of length $I$ crosses exactly $I$ lattice bonds, the current across a fixed bond is
	\[
	J = \frac{I}{L}\,\mathbb{E}_{\hat{\pi}_{N,L}}
	\!\Bigl[\sum_{i=1}^{N}(r_{g_i}-\ell_{g_{i-1}})\Bigr].
	\]
	By periodicity of the ring, $\sum_i\ell_{g_{i-1}}=\sum_i\ell_{g_i}$, so the two sums combine to
	$\sum_i(r_{g_i}-\ell_{g_i})$. By translation invariance of the ring and $N/L=\rho$,
	\[
	\frac{1}{L}\,\mathbb{E}_{\hat{\pi}_{N,L}}
	\!\Bigl[\sum_{i=1}^{N}r_{g_i}\Bigr]
	= \frac{N}{L}\,\mathbb{E}_{\nu_N}[r_g]
	= \rho\,\mathbb{E}_{\nu_N}[r_g],
	\]
	and similarly for $\ell_g$, where $\nu_N$ denotes the single-site marginal of $\hat{\pi}_{N,L}$. This gives \eqref{eq:J_micro}:
	\[
	J = \rho\,I\,\mathbb{E}_{\nu_N}[r_g-\ell_g].
	\]
	By Lemma~\ref{lem:equivalence}, $\nu_N\to\nu_\lambda$ weakly as $N,L\to\infty$ with $N/L\to\rho$, so $\mathbb{E}_{\nu_N}[r_g-\ell_g]\to\mathbb{E}_{\nu_\lambda}[r_g-\ell_g]$. Using \eqref{right_jump}--\eqref{left_jump} and \eqref{eq:nu_lambda_def},
	\begin{align}
		\mathbb{E}_{\nu_\lambda}[r_g-\ell_g]
		&= \frac{1}{Z(\lambda)}
		 \sum_{g> I}(r^\star-\ell^\star)\, e^{J_{g-I}-J_g}
		\cdot e^{J_g-\lambda g}
		\nonumber\\
		&= \frac{r^\star-\ell^\star}{Z(\lambda)}
		 \sum_{g> I} e^{J_{g-I}-\lambda g}.
	\end{align}
	 The substitution $g'=g-I\ge1$ gives, using $J_{g'}=0$ for $g'\le I$,
	\begin{align*}
		 \sum_{g> I}e^{J_{g-I}-\lambda g}
		& = \sum_{g'\ge 1}e^{J_{g'}-\lambda(g'+I)}\\
		& = e^{-\lambda I}\sum_{g'\ge 1}e^{J_{g'}-\lambda g'}\\
		& = e^{-\lambda I}Z(\lambda),
	\end{align*}
	hence
	\begin{align*}
		\mathbb{E}_{\nu_\lambda}[r_g-\ell_g]
		& = \frac{r^\star-\ell^\star}{Z(\lambda)}\cdot e^{-\lambda I}Z(\lambda)\\
		& 	= (r^\star-\ell^\star)\,e^{-\lambda I}.
	\end{align*}
	Inserting into \eqref{eq:J_micro} yields \eqref{eq:current_exact}:
	\[
	J(\rho) = \rho\,I\,(r^\star-\ell^\star)\,e^{-\lambda(\rho)\,I}.
	\]

\end{document}